# Subclass Classification of Gliomas Using MRI Fusion Technique

Kiranmayee Janardhan[1]  · Christy Bobby Thomas[1]

## Abstract

Glioma, the prevalent primary brain tumor, exhibits diverse aggressiveness levels and prognoses. Precise classification of glioma is paramount for treatment planning and predicting prognosis. This study aims to develop an algorithm to fuse the MRI images from T1, T2, T1ce, and fluid-attenuated inversion recovery (FLAIR) sequences to enhance the efficacy of glioma subclass classification as no tumor, necrotic core, peritumoral edema, and enhancing tumor. The MRI images from BraTS datasets were used in this work. The images were pre-processed using max–min normalization to ensure consistency in pixel intensity values across different images. The segmentation of the necrotic core, peritumoral edema, and enhancing tumor was performed on 2D and 3D images separately using UNET architecture. Further, the segmented regions from multimodal MRI images were fused using the weighted averaging technique. Integrating 2D and 3D segmented outputs enhances classification accuracy by capturing detailed features like tumor shape, boundaries, and intensity distribution in slices, while also providing a comprehensive view of spatial extent, shape, texture, and localization within the brain volume. The fused images were used as input to the pre-trained ResNet50 model for glioma subclass classification. The network is trained on 80% and validated on 20% of the data. The proposed method achieved a classification of accuracy of 99.25%, precision of 99.30%, recall of 99.10, F1 score of 99.19%, Intersection Over Union of 84.49%, and specificity of 99.76, which showed a significantly higher performance than existing techniques. These findings emphasize the significance of glioma segmentation and classification in aiding accurate diagnosis.

## Introduction

Glioma, a heterogeneous form of brain tumor, exhibits varying morphological characteristics across its different subclasses/subregions, which include the necrotic core (NCR), peritumoral edema (ED), and enhancing and non-enhancing tumor regions (ET/NET) [1]. It arises from glial cells in the central nervous system, which includes the brain and spinal cord. Gliomas can vary in severity and aggressiveness; some are benign and slow growing, while others are malignant and rapidly growing. Comprehending these subclasses is essential for developing tailored treatment strategies. Research on categorizing gliomas using deep learning and machine learning has generated promising results with 80–90% accuracy rates. Accuracy is limited by existing algorithms' difficulty distinguishing the classes within gliomas. Addressing these gaps [2] is essential for the development of more efficient glioma classification techniques and diagnostic tools. Utilizing BraTS datasets for 2D and 3D data segmentation involves delineating and identifying various regions within brain MRI scans. In this study, the segmentation of gliomas, a crucial aspect of medical image analysis, has been addressed using a modified UNET. This architecture, based on deep convolutional neural networks (CNN), excels at feature extraction and pattern recognition, allowing it to precisely locate tumors by capturing fine details and spatial relationships in the images. This methodology enhances the accuracy and efficiency of glioma segmentation, providing valuable insights for neuro-oncology diagnosis and therapy planning.

The modified UNET architecture is utilized in glioma segmentation due to its ability to capture contextual information, particularly when combined with both 2D and 3D segmentation results [3]. When coupled with ResNet50 in deep learning, it aids in subclass classification (necrotic, enhancing, non-enhancing tumor, and peritumoral edema)

✉ Kiranmayee Janardhan
kiranmayee.j@msruas.ac.in

1    Department of Electronics and Communication Engineering, Ramaiah University of Applied Sciences, Bengaluru, India

of gliomas. This approach, combining UNET segmentation and ResNet50 classification, offers enhanced glioma characterization, promising improved diagnostic precision, and valuable insights for personalized treatment approaches in medical imaging.

In the work [3], Unified Visualization and Classification Network (UniVisNet), a framework that enhanced both classification performance and high-resolution visual explanation generation was developed. By introducing a subregion-based attention mechanism and fusing multiscale feature maps, UniVisNet achieved superior visual explanations without additional steps. Experiments demonstrate its success, with remarkable results on glioma subregion grading with AUC of 94.7%, Accuracy of 89.3%, Sensitivity of 90.4%, and Specificity of 85.3%. In the work [4] a Densenet201 Pre-Trained Deep Learning Model was fine-tuned using imbalanced data deep transfer learning. To extract deep information on tumor types, features from the average pool layer are utilized. Due to insufficient precision, two feature selection techniques were developed: Entropy-Kurtosis-based High Feature Values (EKbHFV) and a modified genetic algorithm (MGA). MGA-selected features are refined by a new threshold function. Both EKbHFV and MGA-based features were then fused using a non-redundant serial-based approach and classified with a multiclass SVM cubic classifier. Experiments on BraTS 2018 and BraTS 2019 datasets achieved over 95% accuracy without data augmentation. In work [5], the method optimizes deep learning features for the four modalities for tumor classification using ResNet101 pre-trained model and transfer learning. To tackle redundant features, differential evaluation and particle swarm optimization find optimal features which are then fused, and PCA is applied for further optimization. The final feature vector is classified using various classifiers, achieving a 25.5×speedup in prediction time with 94.4% accuracy. In the work [6], a hybrid ensemble learning model and feature extraction method is proposed for glioma classification from fused MRI sequences. Combining Discrete Wavelet Decomposition, Central Pixel Neighbourhood Binary Pattern, and Gray Level Run Length Matrix, this approach improved classification accuracy. The eXtreme Gradient Boosting classifier with random forest is used with this hybrid feature extraction method and evaluation on local and global datasets, including BRATS 2013 and BRATS 2015, with various MRI fusion combinations, demonstrated 99.25% accuracy when implemented with T1C+T2+Flair MRI sequences.

In the work [7], a domain knowledge-modified CNN architecture and Stack Encoder-Decoder network are combined with an evolutionary optimization algorithm for hyperparameter selection. The improved Grey Wolf algorithm with updated Jaya algorithm criteria enhances learning speed and accuracy. A novel parallel pooling approach fuses selected features, which are then classified using machine learning and neural networks. Experiments on BraTS 2020 and BraTS 2021 datasets yield an improved average accuracy of 98% and a maximum single-classifier accuracy of 99%. In the work [8], a novel multistream deep CNN architecture is proposed for glioma grading, addressing issues of brain tumor classification from multi-sensor images. The architecture extracts and fuses features from T1-MRI, T2-MRI, and FLAIR sensors, leveraging their unique contrast sensitivities. Key contributions include the multistream deep CNN architecture, sensor fusion for feature aggregation, and 2D image slice augmentation to mitigate overfitting. Experiments on two datasets show promising results, with test accuracies of 90.87% and 89.39% for classifying low/high-grade gliomas and gliomas with/without 1p19q codeletion, respectively. In this work [9], the research explores multiclass brain tumor classification using Deep Learning and Machine Learning techniques. Utilizing Convolutional Neural Networks (AlexNet, ResNet-18, and GoogLeNet) for feature fusion and SVM and KNN for classification, the proposed method achieved 98% accuracy on 15,320 MRI images, showcasing the potential of these advanced algorithms to improve brain tumor diagnosis and patient outcomes. In the study [10], the researchers evaluated fine-tuned, pretrained deep CNN models for brain tumor classification in MRI images. VGG16, VGG19, ResNet50, ResNet101, and InceptionResNetV2 achieved state-of-the-art accuracy and all models surpassed 99% accuracy on subclass MRI classification.

The study [11] introduces two fast and effective brain tumor detection techniques using deep convolutional neural networks (CNNs) with MRI data. Figshare and BraTS 2018 datasets were utilized and conditional random fields were applied to refine segmentation. The first CNN architecture classifies tumors into gliomas, meningiomas, or pituitary tumors, while the second differentiates between high- and low-grade gliomas. Intensity normalization and data augmentation are also explored to enhance detection. The first CNN achieved 97.3% accuracy and 95.8% DSC, while the second achieved 96.5% accuracy and 94.3% DSC. In the study [12] presents a convolutional neural network based on complex networks (CNNBCN) with a modified activation function for MRI-based tumor classification. Unlike manually designed networks, CNNBCN uses randomly generated graphs mapped into a neural network. The modified CNNBCN model achieves a 95.49% classification accuracy, surpassing several existing models. It also shows lower test loss than ResNet, DenseNet, and MobileNet. This approach not only improves classification performance but also advances neural network design methodology. In this study [13], the researchers have developed a system for classifying MRI brain tumor images into four categories using VGG-16, ResNet-50, and AlexNet models in an ensemble approach. The ensemble model achieved a 99.16% accuracy, outperforming other methods like Naive

Bayes, decision trees, random forests, and DNNs, without postprocessing. In this study [14], an automated method for brain tumor detection using MRI is proposed. MRI images are first pre-processed to enhance quality. Two pre-trained deep learning models extract features, which are then combined into a hybrid vector using partial least squares (PLS). Top tumor locations are identified through agglomerative clustering, and features are classified using a head network. This method achieves a classification accuracy of 98.95%, surpassing existing approaches.

In this research paper, the Introduction is discussed, followed by the Methodology with dataset details, segmentation and classification and implementation details. The Results section also consists of a comparative analysis with the works from literature. This is followed by the Conclusion section.

## Methodology

This study utilizes deep learning methods to implement segmentation, fusion, and classification of gliomas. The workflow for the proposed model is illustrated in Fig. 1.

### Dataset and Preprocessing

This study utilizes the BraTS datasets from 2018, 2019, and 2020 [15]. The datasets encompass cases of Glioblastoma/

High-Grade Glioma (GBM/HGG) and Lower Grade Glioma (LGG). In BraTS 2018, there were 285 cases (210 HGG, 75 LGG), in BraTS 2019, 335 cases (259 HGG, 76 LGG), and BraTS 2020, 369 cases (293 HGG, 76 LGG). The MRI images consist of four modalities: T1-weighted (T1), T2-weighted (T2), post-contrast T1-weighted (T1ce), and fluid-attenuated inversion recovery (FLAIR). These MRI images were obtained from multi-parametric MRI scans conducted in routine clinics, using MRI scanners ranging from 1 to 3 T from 19 multi-center institutions, and stored in the Neuroimaging Informatics Technology Initiative (NIfTI) format. All four MRI modalities underwent preprocessing steps including bias field correction, skull stripping, and co-registration into the same anatomical structure template [16].

In the preprocessing pipeline, the original MRI volumes have dimensions of $240 \times 240 \times 155$, with 155 slices representing the depth of the scan. Each of these volumes contains manually segmented scans. The goal of preprocessing is to optimize these inputs for neural network training. To achieve this, Min–Max Scaling is applied, normalizing the pixel values between 0 and 1, which aids in more stable and consistent gradient updates during training. This normalization step is essential for smooth convergence, preventing the model from becoming biased due to variations in pixel intensity across different scans. Additionally, to align with the label schema of the BraTS dataset, class '4' labels were reassigned to '3', as label '3' does not exist in the dataset.

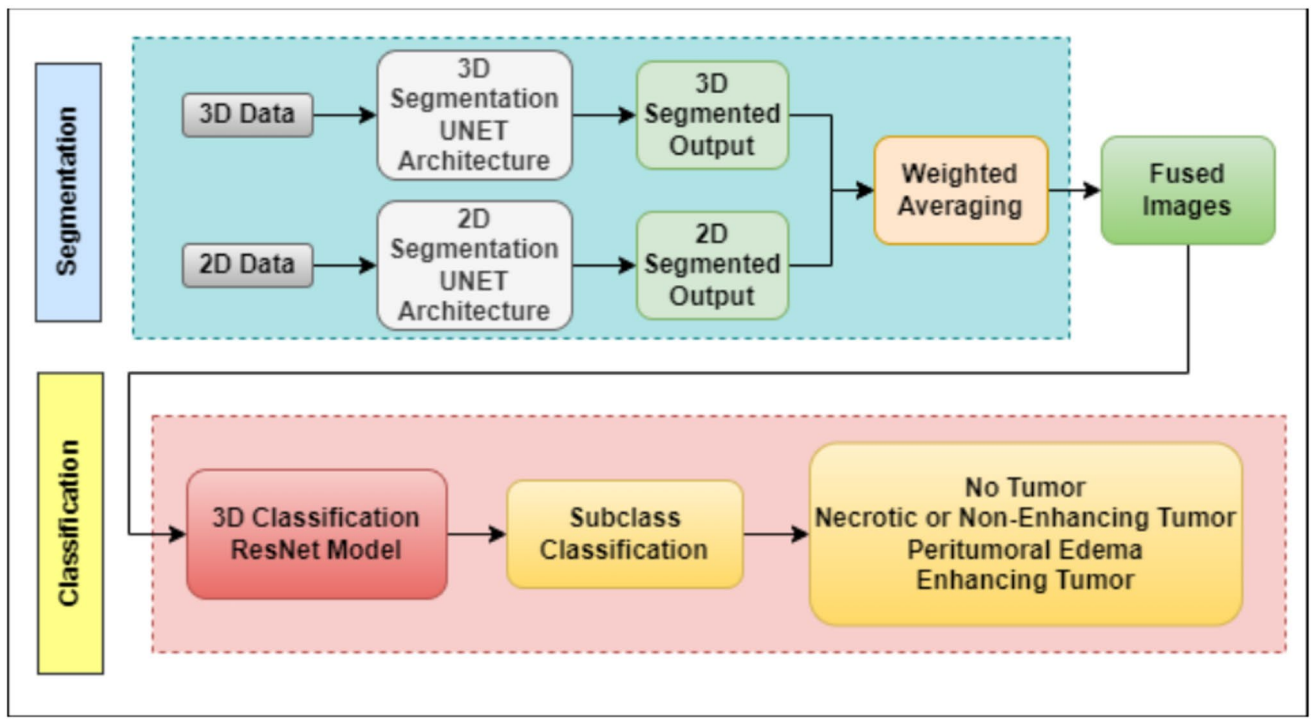

**Fig. 1** Workflow of glioma segmentation and subclass classification

**Algorithm** The comprehensive algorithm for the automated segmentation and classification is given below.

---

**Comprehensive Algorithm for Automated Segmentation and Classification of Gliomas**

I. Data Preparation
1. Image Acquisition and Resizing:
   - Acquire multi-sequence MRI scans: FLAIR, T1, T1CE, and T2.
   - Crop all images to 128x128 pixels using:
     Cropped Image = Crop (Original Image, $(128 \times 128)$)
2. Dataset Utilization:
   - Use BraTS datasets from 2018, 2019, and 2020.
   - Split datasets into 80% for training and 20% for validation.

II. Segmentation Using Modified UNET
1. Down-sampling Path:
   - Apply convolution and max pooling iteratively:
     $$C_l = f(W_l \cdot I_l + b_l), \qquad P_l = MaxPool(C_l)$$
   - $C_l$: Output of convolutional layer $l$.
   - $P_l$: Output of pooling layer $l$.
2. Up-sampling Path:
   - Perform upsampling and concatenate with corresponding feature maps from the down-sampling path:
     $$U_l = U_p(C_l), \qquad c_{l+1} = Concat(U_l, c_{d-1})$$
   - Apply convolution to up-sampled feature maps:
     $$C_l' = f(W_l' * C_{l+1} + b_l')$$
3. Output Layer:
   - Generate segmented output using final convolution:
     $$O = f(W_f \cdot C_n + b_f)$$

III. Fusion of 2D and 3D Data
1. Weighted Averaging:
   - Fuse 2D and 3D segmented outputs using:
     $$S_{fused} = \alpha \cdot S_{2D} + (1 - \alpha) \cdot S_{3D}$$

IV. Classification Using Modified ResNet50
1. Feature Extraction:
   - Pass fused segmented images through modified ResNet50:
     Feature Maps = ResNet50$(S_{fused})$
2. Softmax Classification:
   - Convert logits to probabilities for subclass prediction:
     $$f_i(z) = \frac{e^{z_i}}{\sum_{j=1}^{K} e^{z_j}}$$
3. Classification Decision:
   - The model selects the class with the highest softmax probability.

V. Performance Metrics
1. Dice Coefficient Calculation:
   - Compute Dice similarity for segmentation:
     $$D_{Coe} = \frac{2 \times I_n + S_m}{T_{pp} + P_{pp} + S_m}$$
   - Calculate Total Dice Coefficient for subclass accuracy:
     $$T_D = \frac{1}{C_{num}} \sum_{i=1}^{C_{num}} \frac{2 \times I_n + S_m}{T_{pp} + P_{pp} + S_m}$$

Notations:
- $I_l$: Input to the $l$-th layer.
- $W_l$, $b_l$: Weights and biases for the $l$-th convolutional layer.
- $C_l$: Output of the $l$-th convolutional layer.
- $P_l$: Output of the $l$-th pooling layer (max pooling).
- $U_l$: Upsampled output of the $l$-th layer.
- $C_l'$: Convolution output of upsampled layer $l$.
- $C_{l-1}$: Concatenated output from the upsampling path and corresponding features from the down-sampling path.
- $W_l'$, $b_l'$: Weights and biases for the convolutional layers in the upsampling path.
- $W_f$, $b_f$: Weights and biases for the final convolutional output layer.
- $O$: Final output of the UNET (segmented image).
- $S_{2D}$, $S_{3D}$: 2D and 3D segmented outputs, respectively.
- $S_{fused}$: Fused segmented output using weighted averaging.
- $\alpha$: Weighting factor for fusing 2D and 3D data.
- $z$, $z_i$: Logits vector and the i$^{th}$ logit for each class.
- $K$: Total number of classes (subclasses).
- $f_i(z)$: Softmax probability of the i$^{th}$ class.
- $D_{Coe}$: Dice coefficient for measuring overlap between predicted and ground truth segmentation.
- $T_{pp}$, $P_{pp}$: Predicted and ground truth segmentations.
- $S_m$: Smoothing term added to prevent division by zero in Dice coefficient calculation.
- $T_D$: Total Dice Coefficient across all subclasses.
- $I_n$: Intersection of predicted and ground truth segmentations.
- $C_{num}$: Number of classes for subclass classification  here  $C_{num} = 4$

---

By focusing on the ROI and minimizing irrelevant background noise, cropping to $128 \times 128$ pixels preserves essential features and tumor details, allowing the model to concentrate on high-resolution details within the relevant area for more accurate segmentation. Retaining only the necessary region also supports continuity and detail in 3D space, helping the model maintain spatial accuracy across slices, which is crucial for capturing depth and density variations along tumor boundaries. This targeted cropping further streamlines processing, enabling the model to allocate resources more efficiently toward significant features, enhancing segmentation precision and overall model performance without compromising spatial resolution within the ROI. This approach provides an effective balance between computational efficiency and the ability to capture critical details, making it well-suited for tasks requiring high-resolution focus on specific areas, which are essential for accurate classification and segmentation. Subsequently, the data is partitioned into training (80%) and validation (20%) datasets.

## Segmentation of 2D and 3D with UNET

### Architecture

Figure 2 shows the modified UNET architecture proposed in this work by modifying the feature map's dimension for accurate and automatic glioma segmentation. UNET, a fully convolutional neural network designed for biomedical image segmentation, played a pivotal role in this context due to its unique architecture and capabilities [17]. The convolution nature enabled it to capture both local and global contextual information from the MRI image which is essential for accurately segmenting glioma subclasses. The skip connections bridge the semantic gap between low-level features (edges and textures) and high-level features (shapes and context), which allowed the model to capture the heterogeneity of different tumor subclasses effectively. Glioma subclasses exhibit different characteristics in different MRI modalities and UNET has been trained separately for each modality. 2D segmentation captured slice-wise information, whereas 3D segmentation captured volumetric information, which is vital for understanding the spatial relationships between different glioma subclasses.

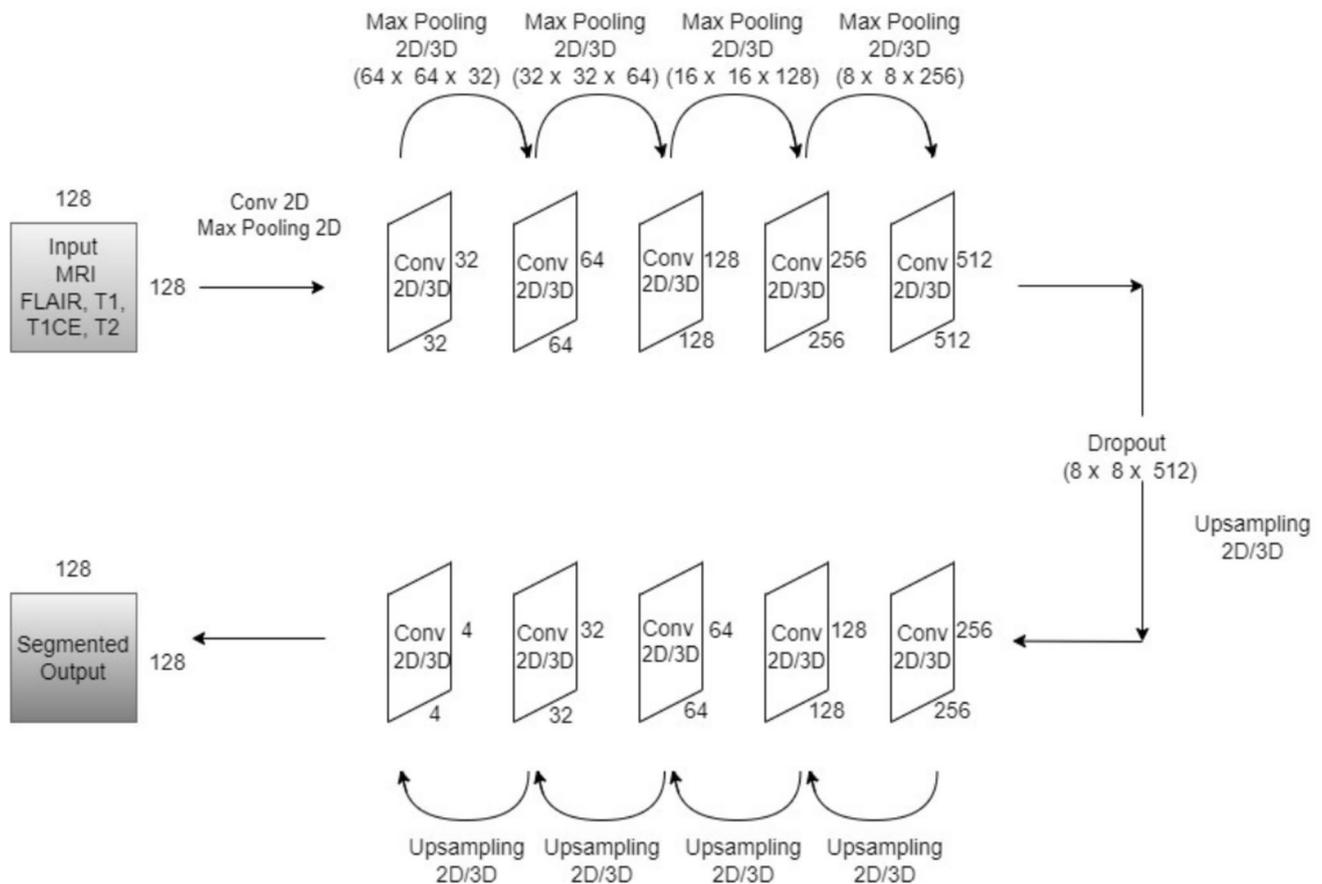

**Fig. 2** Modified UNET Architecture used for Segmentation of Gliomas

In the BraTS datasets, the training set consists of samples with four ground truth labels corresponding to four distinct regions: background (label 0), necrotic and non-enhanced tumor (label 1), peritumoral edema (label 2), and enhanced tumor (label 4). To streamline the analysis, the non-zero labels were consolidated into three combined subclasses, which include enhanced tumor (ET: label 4), tumor core (TC: union of labels 1 and 4), and whole tumor (WT: union of labels 1, 2, and 4). The WT, TC, and ET regions of the MRI images were then derived by utilizing a pre-trained segmentation model.

Given that gliomas are characterized by their growth within the brain tissue and their tendency to intermingle with normal brain tissues, the assessment of the surrounding area is crucial. Consequently, the whole tumor regions were utilized as the ROIs for segmentation purposes. To ensure consistency in image dimensions, the original MRI images were centrally cropped to a size of $128 \times 128$. Furthermore, min–max normalization was applied to the cropped images to enhance the homogeneity of their intensities, thereby facilitating more accurate and efficient image analysis. The dense blocks within the network architecture enabled the extraction of image features. Subsequently, these features were mapped to three glioma subclasses using fully connected layers. Specifically, in 3D segmentation, to preserve the inherent 3D structural features of MRI images, they were directly input to the model without converting them to 2D slices.

During the training phase, various parameters were configured to optimize the model's performance. The imbalanced nature of datasets, particularly in subclass classifications, poses a significant challenge as it can lead to biased predictions favouring the majority class. To mitigate this, focal loss was implemented during training, which is specifically designed to down-weight the loss contribution from well-classified examples and focus more on hard-to-classify instances. This approach helped to ensure that the model paid sufficient attention to minority classes, which is crucial for achieving reliable subclass classifications. In this work, the focal loss was integrated into the training process alongside the ResNet50 architecture, allowing the model to learn effectively from the fused 2D and 3D segmentation outputs. The performance metrics, such as precision, recall, and F1-score, were monitored to evaluate the impact of this technique on classification accuracy across all classes. The results demonstrated an improvement in the model's ability to classify underrepresented subclasses, reinforcing the importance of this approach in handling imbalanced datasets.

The initial learning rate was set to $5 \times 10^{-4}$ with a scheduler optimization updating strategy. The Adam algorithm with impulse was used as the optimizer, the batch size was set to 8, and the number of training epochs was set to 100. Multiple strategies were utilized to overcome overfitting during the training process. These strategies included sample normalization, data augmentation, applying L2 normalization to model loss, designing a dropout layer for the model, and setting weight decay for the optimizer. The 2D and 3D segmentation models were implemented in Google Colab Pro＋. The Medical Open Network for AI (MONAI) toolkit [18] was used to implement various data augmentation techniques, such as dimension resizing, random rotation, random scaling, random Gaussian noise addition, and random contrast adjustment. To reduce the computational cost associated with a large number of model parameters, an NVIDIA Tesla A100 GPU was utilized to decrease the running time for segmentation model training and validation.

## Fusion of 2D and 3D Segmented Data

The weighted averaging of 2D and 3D segmented MRI images involved the integration of segmentation masks obtained from various imaging modalities [18]. The primary rationale for using a feature fusion-based approach is to generate informative and distinctive features from MRIs, as these features play a pivotal role in ensuring precise tumor classification. By combining data from multiple sources, this technique enhances the diagnostic capabilities of the system, ultimately leading to more accurate identification and categorization of tumors. To determine the significance of each modality's contribution based on its relevance or information content, a weighted averaging strategy was used as expressed in Eq. (1), incorporated segmentation values for the same voxel or pixel from both 2D $S_{2D}$ and 3D $S_{3D}$ segmentations. This formula enables a sophisticated fusion of data from various modalities, emphasizing the importance of each modality's unique strengths in outlining tumor regions through their weighted contributions. This strategy involved computing a weighted average for every pixel or voxel in the segmentation masks, considering the corresponding values from both 2D and 3D segmentations. The 2D segmentations capture better sharp boundary transitions, while 3D segmentation provides insights into the tumor's spatial extent [19]. By integrating both through a weighted average, this method compensates for any potential loss of fine detail due to image cropping and ensures a consistent delineation of tumor regions. By capitalizing on the unique strengths of different imaging perspectives, this approach demonstrated its effectiveness in delivering a more reliable and comprehensive subclass classification of gliomas. This fusion of 2D and 3D segmentations resulted in a more robust identification of tumor subclasses, ultimately improving diagnostic accuracy and treatment planning.

$$S_{fused} = \alpha . S_{2D} + (1 - \alpha) . S_{3D} \tag{1}$$

In this context, $\alpha$ represents the weight given to the 2D segmentation, while $(1 - \alpha)$ corresponds to the weight attributed to the 3D segmentation. The weighting factor $(\alpha)$ in the fusion process plays a crucial role in balancing the contributions of 2D and 3D segmentations in the overall classification. The value of $\alpha$ was determined based on the importance of capturing both boundary precision and volumetric consistency. 2D segmentation is particularly effective at capturing fine boundary transitions, while 3D segmentation provides a more holistic view of the tumor's spatial structure. Hence, the weighting strategy was designed to emphasize the strengths of each modality. To determine the optimal value for $\alpha$, an experimental tuning process was conducted. Different values for $\alpha$ were tested to assess their impact on model accuracy during training. A grid search method was used to optimize this parameter by evaluating its performance on the validation set. The chosen value of $\alpha$ reflected the best trade-off between boundary accuracy from 2D segmentations and spatial context from 3D segmentations, with the aim of improving overall glioma classification accuracy.

An approximate value for $\alpha$, found to offer the best performance in this study, was 0.6. This value gives slightly more weight to the 2D segmentation due to its ability to preserve boundary details, while still maintaining a significant contribution from the 3D segmentation to account for tumor depth and spatial features. Although $\alpha$ was not directly optimized through automated processes like hyperparameter tuning, its selection was crucial in ensuring the fusion technique's effectiveness in accurately delineating tumor regions and improving classification outcomes. The primary goal of using a weighted averaging strategy was to harness the unique advantages of various imaging modalities, thereby generating a more precise and robust depiction of glioma boundaries. By capitalizing on the individual strengths of each modality, this approach aimed to enhance clinical decision-making and treatment planning for patients affected by gliomas.

The novelty of the proposed fusion technique lies in its integration of 2D and 3D segmentations, extending beyond standard segmentation with a modified UNET and classification with ResNet50. The innovation is centered around the weighted averaging strategy, which combines the complementary strengths of 2D and 3D segmentation outputs. While 2D segmentation excels at capturing sharp boundary transitions, 3D segmentation offers a more complete view of the tumor's spatial extent. By assigning different weights to these segmentations, the fusion process emphasizes significant features from both perspectives, compensating for any loss of fine detail during image cropping. This weighted fusion preserves both boundary precision and volumetric consistency, resulting in more robust and accurate glioma classification, ultimately improving diagnostic precision and facilitating better clinical decision-making.

ResNet50's architecture is particularly well-suited for subclass glioma classification when using inputs from the fused 2D and 3D segmentations. Its deep residual learning framework excels at learning complex hierarchical features, overcoming vanishing gradient issues common in deep networks. This capability is crucial for handling the fused segmentation data, where the weighted averaging process integrates fine 2D boundary details with 3D spatial context. ResNet50 effectively captures multidimensional features through its residual connections, refining and enhancing the fused outputs layer by layer. Its convolutional layers detect subtle patterns such as texture variations, shape irregularities, and volumetric nuances, leading to precise subclassification of gliomas. This synergy between the weighted fusion process and ResNet50's deep learning capabilities results in excellent classification accuracy and improves overall diagnostic precision.

## Deep Transfer Learning for Classification using ResNet50

Deep neural networks are capable of learning discriminative features from imaging data by integrating multiple convolutional layers. Various deep network models and their variations have demonstrated varying levels of classification performance. ResNet50 has showcased superior performance due to its ability to enhance feature propagation from one block to the next. This is achieved by addressing the vanishing gradient problem, which is a common challenge faced in training deep neural networks [20]. Figure 3 shows the deep transfer learning of ResNet implementation for the classification of gliomas.

In the proposed approach, fused images resulting from the combination of 2D, and 3D segmentation outputs serve as input to the ResNet50 model. Figure 4 shows the implementation model for the classification of gliomas. The training process focuses on segmented ROIs for subclass classification: No Tumor, Necrotic or Non-Enhancing Tumor Core, Peritumoral edema, and Enhancing Tumor. The ResNet50 encoder, specifically designed for feature extraction from the segmented glioma region, undergoes training modifications. Softmax layer activation is applied during the classification process, generating a probability distribution based on raw class scores or logits. The assigned probabilities represent the likelihood of each segmented subclass. The model's classification decision, based on the softmax probabilities, selects the class with the highest probability as the final prediction, as expressed by the Eq. (2). The output vector of the softmax layer gives the classification probabilities for each segmented

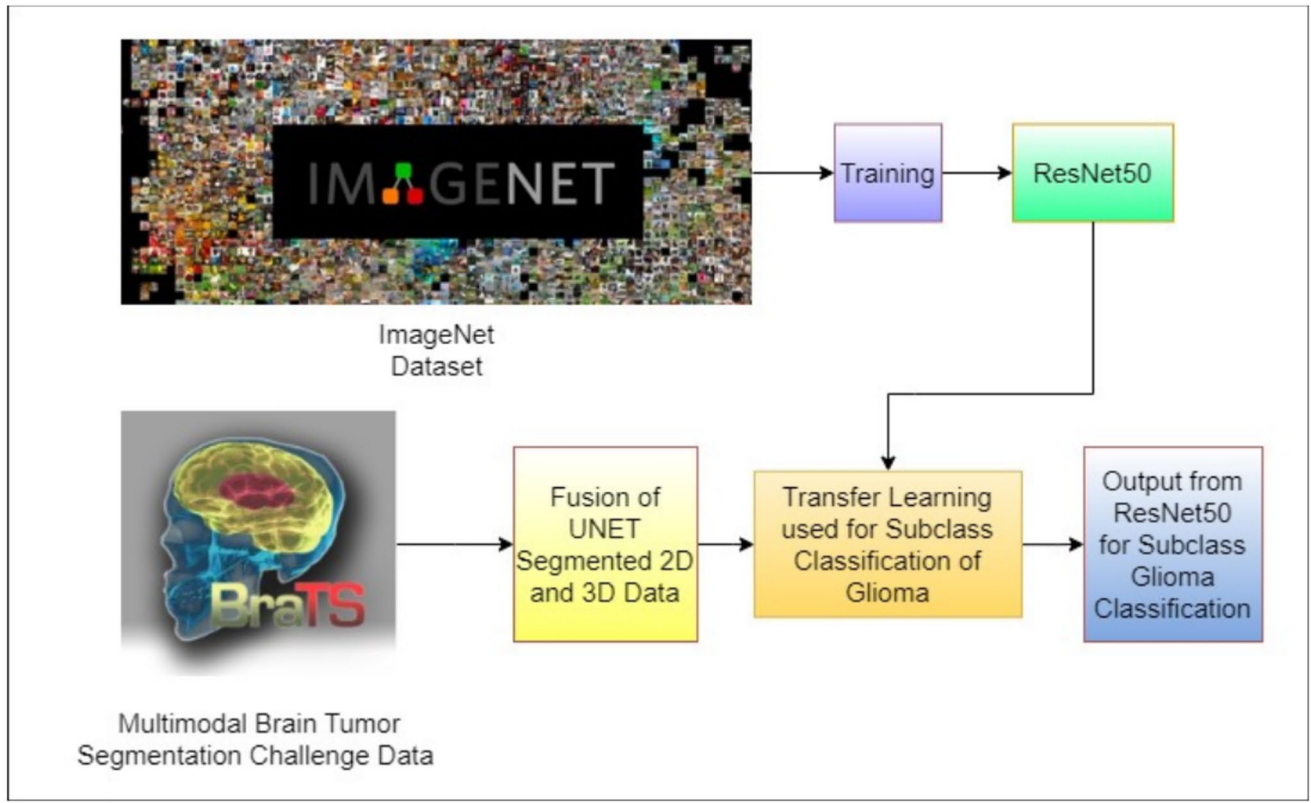

**Fig. 3**  Transfer Learning of ResNet implementation for Classification of Gliomas using ImageNet

subclass, forming the basis for the model's decision-making process.

$$f_i(z) = \frac{e^{z_i}}{\sum_{j=1}^{K} e^{z_j}} \text{ for } i = 1, \ldots, K \text{ and,} \qquad (2)$$

$$z = (z_1, \ldots, K) \in \mathbb{R}^K$$

where, $f_i$ to mean the $i$ th element of the vector of class scores $f$. The Dice Coefficient ($D_{CoE}$) loss equation is given by Eq. (3):

$$D_{CoE} = \frac{2 \times I_n + S_m}{T_{pp} + P_{pp} + S_m} \qquad (3)$$

where, $I_n$ stands for Intersection and $S_m$ stands for Smooth, $T_{pp}$ is the true positive pixels and $P_{pp}$ is the predicted positive pixels. The process involves calculating the $D_{CoE}$ for each class and subsequently averaging these coefficients to obtain the Total Dice Coefficient ($T_D$). To avoid potential division by zero issues, a smoothing parameter is introduced.

This parameter ensures a small, non-zero value is added to the numerator and denominator of the Dice coefficient formula for each class, facilitating a stable and meaningful computation of segmentation accuracy across all classes as given by Eqs. (3) and (4):

$$T_D = \frac{1}{C_{num}} \sum_{i=1}^{C_{num}} \frac{2 \times In + Sm}{T_{pp} + P_{pp} + Sm} \text{and} \qquad (4)$$

$C_{num} = 4$ for Subclass Classification

Table 1 provides detailed implementation layer specifications for the ResNet50 model used in glioma subclass classification. The model specifications include the use of TensorFlow version 2.8.2 as the software framework. The architecture employed is ResNet50, which was pretrained on the ImageNet dataset. The number of epochs is set to 100 for BraTS 2018 and BraTS 2020 and 50 for BraTS 2019, for subclass classification task. The weight decay is implemented at a rate of 1e-05 with Cosine Annealing. The batch size was set to 32, as it offered a balance between frequent weight updates, which enhance stability, and efficient training times. This choice allowed the model to process more images simultaneously, potentially speeding up convergence while maintaining effective learning

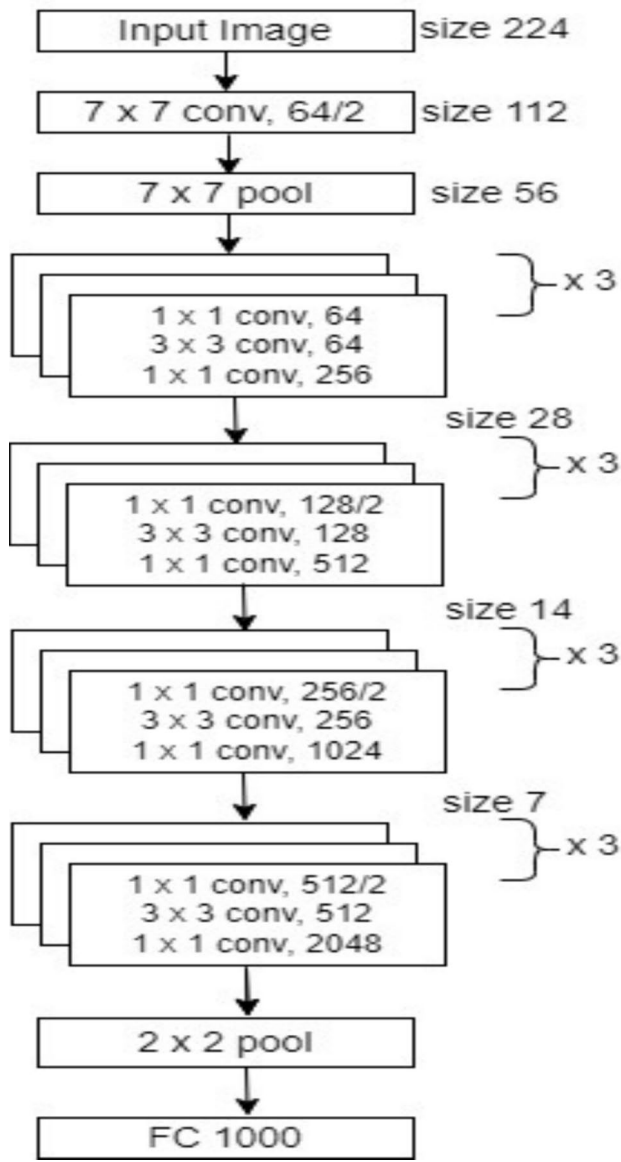

**Fig. 4** Modified ResNet50 implementation for Classification of Gliomas

**Table 1** Modified ResNet50 implementation for Classification of Gliomas

| Layer Name | Output size | 50—Layer |
|---|---|---|
| Conv1_X | 112×112 | 7×7, 64, stride 2 |
| | | 3×3 max pool, stride 2 |
| Conv2_X | 56×56 | $\begin{bmatrix} 1\times1, & 64 \\ 3\times3, & 64 \\ 1\times1, & 256 \end{bmatrix} \times 3$ |
| Conv3_X | 28×28 | $\begin{bmatrix} 1\times1, & 128 \\ 3\times3, & 128 \\ 1\times1, & 512 \end{bmatrix} \times 4$ |
| Conv4_X | 14×14 | $\begin{bmatrix} 1\times1, & 256 \\ 3\times3, & 256 \\ 1\times1, & 1024 \end{bmatrix} \times 6$ |
| Conv5_X | 7×7 | $\begin{bmatrix} 1\times1, & 512 \\ 3\times3, & 512 \\ 1\times1, & 2048 \end{bmatrix} \times 3$ |
| Softmax | 1×1 | Average pool, 1000 FC |

Adam optimizer helped the model learn effectively from the complex data it encountered. These hyperparameters were strategically selected to maximize the model's performance, thereby ensuring stability, efficiency, and accuracy in classifying glioma subclasses.

Segmentation evaluation metrics gauge the efficacy of algorithms in demarcating regions of interest. The Dice Coefficient, also known as the Dice Similarity Coefficient, quantifies the overlap between predicted and true segmentations, with values ranging from 0 (no overlap) to 1 (perfect overlap). This metric assesses the algorithm's performance in accurately identifying and segmenting regions. Accuracy serves as a measurement of the segmentation model's overall predictive correctness. By evaluating the model's ability to accurately identify and classify various regions or objects within an image or volume, the accuracy metric provides a comprehensive appraisal of the model's performance. Hausdorff distance is another commonly used metric to assess the dissimilarity between predicted segmentation (represented as a set of points) and ground truth segmentation. This measure helps to quantify the extent to which the predicted segmentation deviates from the true segmentation, providing valuable insights into the algorithm's segmentation capabilities.

## Implementation

The software implementation for segmentation, fusion, and classification tasks is executed on Google Colab Pro Plus, leveraging the processing capabilities of an NVIDIA V100 graphics processing unit. With a 52 GB RAM allocation, the system is well-equipped to handle the resource requirements dynamics. The learning rate was set to 1e-03, which facilitated smooth weight updates, preventing both rapid convergence to suboptimal solutions and unnecessarily prolonged training durations. This learning rate provided a good balance between training speed and accuracy.

For optimization, the Adam Optimizer is used due to its adaptive learning rate and efficacy in handling sparse gradients. This optimizer integrated the advantages of both AdaGrad and RMSProp, making it particularly suitable for BraTS datasets and high-dimensional parameter spaces, which are characteristic of deep learning tasks. By ensuring faster convergence and enhanced performance, for efficient and seamless execution of the segmentation, fusion, and classification processes. The utilization of these advanced hardware components and software platforms ensures optimal performance and facilitates the efficient completion of the tasks involved in the overall process.

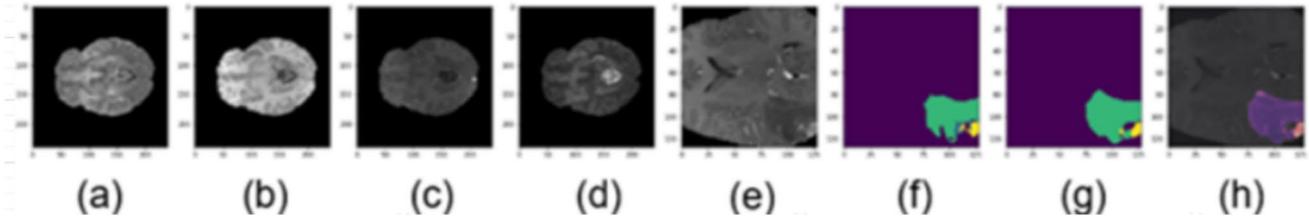

**Fig. 5** **a** FLAIR sequence **b** T1 sequence **c** T1CE sequence **d** T2 sequence **e** Cropped image **f** BraTS Mask **g** Segmented Output **h** 2D and 3D Fused Image from 2020 dataset

**Table 2** Performance metrics with UNET

| BraTS datasets | Segmentation evaluation metrics | | |
|---|---|---|---|
| | DICE | Accuracy | Hausdorff Distance |
| 2018 | 96.36% | 96.93% | 7.6138 |
| 2019 | 96.24% | 96.71% | 6.9283 |
| 2020 | 96.24% | 97.01% | 7.3626 |

## Results

### Visualization of Fusion of 2D and 3D UNET Segmentation

The fusion of multiple imaging modalities in 2D and 3D, such as T1-weighted, T2-weighted, and contrast-enhanced sequences, provides a more comprehensive view of gliomas. This integration enhances the characterization of gliomas by capturing detailed information about the tumor's type, grade, and specific attributes. The multimodal approach offers a more thorough understanding of the tumor's characteristics, as illustrated in Fig. 5, which displays the output of the fused 2D and 3D BraTS datasets. These fused images were given as input to the classification section as seen in the workflow (Fig. 1). Figure 5(a) shows the original FLAIR sequence of the MRI, Fig. 5(b) shows the original T1 sequence, Fig. 5(c) shows the original T1 contrast enhanced sequence, Fig. 5(d) original T2 sequence, Fig. 5(e) shows the cropped image explained in the data preprocessing section, Fig. 5(f) shows the BraTS Mask, Fig. 5(g) shows the segmented output and Fig. 5(h) shows the 2D and 3D Fused Image from BraTS 2020 dataset.

### Segmentation Evaluation Metrics of 2D and 3D Fused Data

The evaluation metrics for the UNET model are presented in Table 2. The modified UNET model achieved an accuracy of 97.01% for the BraTS 2020 dataset, 96.93% for BraTS 2018, and 96.71% for BraTS 2019. These results underscore the potential benefits of model modification and demonstrate the efficacy of the UNET approach in capturing detailed features for precise segmentation. The enhanced accuracy and reliability of glioma classification resulting from the fused approach significantly contribute to improved diagnostic capabilities and potential clinical outcomes.

**Table 3** Subclass classification results with ResNet50

| BraTS Datasets | Training (in %) | | | | | |
|---|---|---|---|---|---|---|
| | Accuracy | Precision | Recall | F1 Score | IOU | Specificity |
| 2018 | 99.24 | 99.36 | 99.03 | 99.19 | 82.59 | 99.78 |
| 2019 | 99.43 | 99.47 | 99.28 | 99.37 | 84.91 | 99.82 |
| 2020 | 99.33 | 99.41 | 99.15 | 99.27 | 66.51 | 99.80 |
| BraTS Datasets | Validation (in %) | | | | | |
| | Accuracy | Precision | Recall | F1 Score | IOU | Specificity |
| 2018 | 99.06 | 99.19 | 98.88 | 99.03 | 82.51 | 99.73 |
| 2019 | 99.25 | 99.30 | 99.10 | 99.19 | 84.49 | 99.76 |
| 2020 | 99.23 | 99.32 | 99.07 | 99.19 | 66.43 | 99.77 |

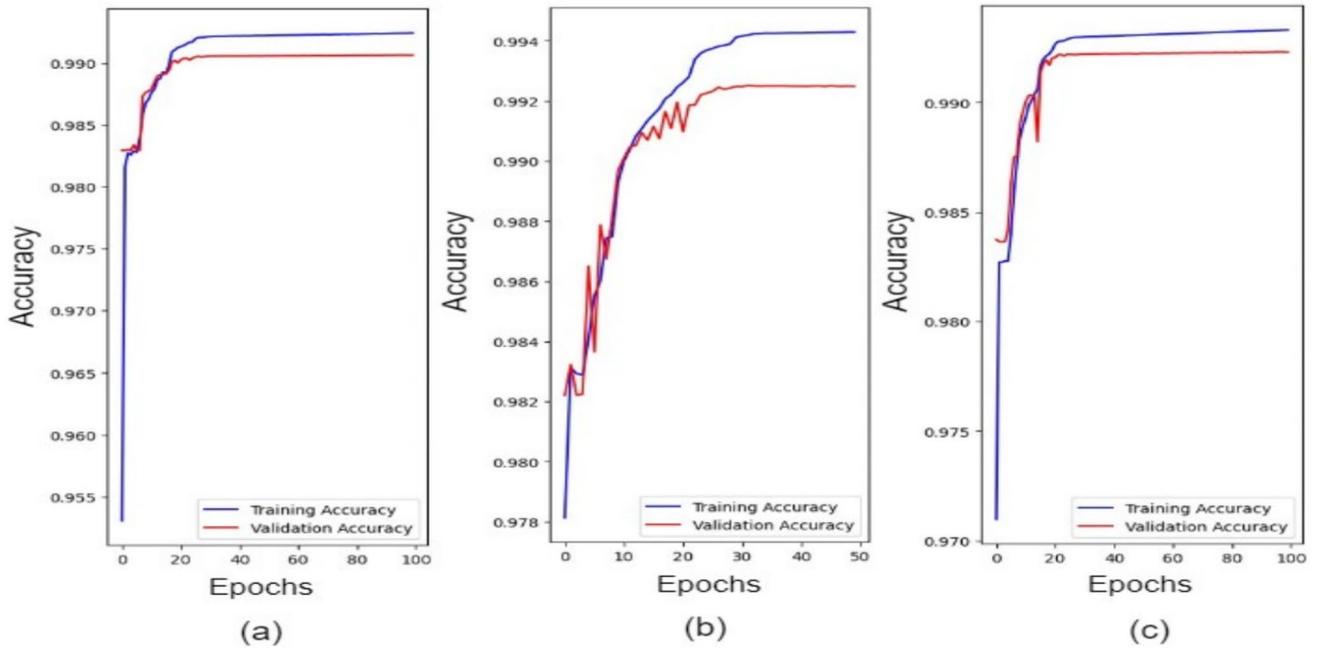

**Fig. 6** Subclass glioma classification accuracy curves **a** BraTS 2018 **b** BraTS 2019 **c** BraTS 2020

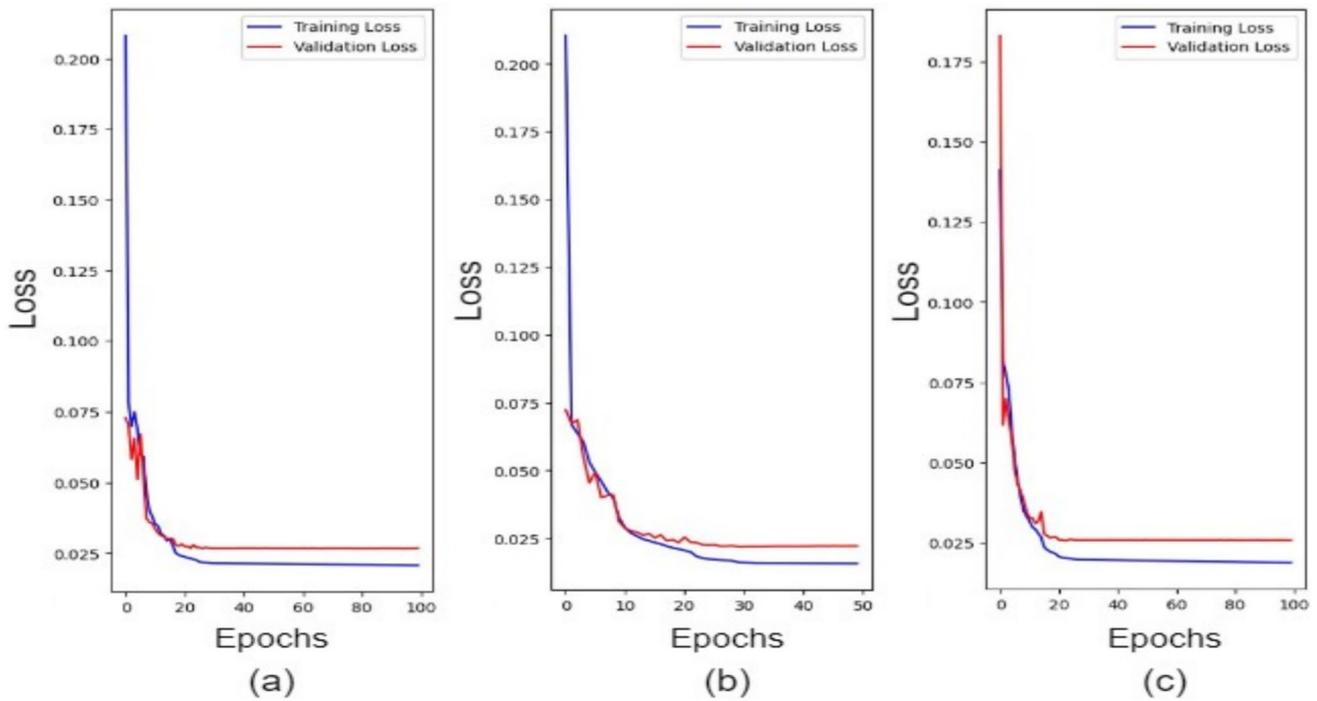

**Fig. 7** Subclass glioma classification loss curves **a** BraTS 2018 **b** BraTS 2019 **c** BraTS 2020

## Subclass Classification Evaluation Metrics

Table 3 shows the performance metrics for ResNet50 subclass classifier outputs during training and validation.

Figure 6 and Fig. 7 show the subclass glioma classification accuracy curves. Figure 6(a), Fig. 6(b), and Fig. 6(c) show the accuracy curves and Fig. 7(a), Fig. 7(b), and Fig. 7(c) show the loss curves for BraTS 2018, 2019, and 2020 respectively.

## Statistical Significance Analyses

Using GraphPad Prism software version 10.3.1, a One-Way ANOVA was performed to assess the statistical significance of differences in training and validation metrics across the years 2018, 2019, and 2020. The analysis revealed that for most metrics—including Accuracy, Precision, Recall, F1 Score, and Specificity—the p-values exceeded the threshold of 0.05, indicating no significant variation between the years. However, the IOU (Intersection over Union) metric for 2020 presented a p-value less than 0.05, suggesting a difference compared to the earlier years. This suggests that while most performance metrics remained stable over time, the IOU showed variability in 2020. To further support the findings, a Kruskal–Wallis test was conducted to account for any potential non-normal distribution in the data. The results of the Kruskal–Wallis test were consistent with those of the ANOVA, reinforcing the conclusion that there were no significant differences in Accuracy, Precision, Recall, F1 Score, and Specificity across the years. The IOU metric's variability in 2020 was once again highlighted. Overall, both statistical tests confirmed the stability of the majority of the performance metrics. This shows that the performance gains are not due to random chance or overfitting of the model.

## Comparative Analysis – Classification Performance Evaluation

Table 4 shows the comparative analysis of the classification results of the performances with the work from literature for subclass classification. In the work by Prasetyo et al., showed that they achieved 99% accuracy using various ensemble methods [10]. The work by Zahid et al. used SWO and ResNet101 for an accuracy of 94.40% [5]. In the work by Zheng et al., showed that they achieved 89.3% accuracy using the attention mechanism and fusing multiscale feature maps [3]. The work by Sharif et al. used a fusion of EKbHFV and MGA-based features and classified using a cubic SVM classifier to obtain an accuracy of 95% [4]. The proposed fusion method achieves a remarkable 99.25% subclass classification accuracy by combining 2D and 3D glioma images with ResNet's pre-trained ImageNet model. This innovative approach capitalizes on the complementary strengths of 2D and 3D MRI data, providing a comprehensive understanding and analysis of gliomas.

The proposed fusion method achieves an impressive 99.25% accuracy in subclass classification by integrating 2D and 3D glioma images using ResNet50's pre-trained ImageNet model. This innovative approach leverages the complementary strengths of both imaging modalities, offering a more comprehensive understanding and analysis of gliomas. The 3D volumetric representation captures intricate spatial relationships among anatomical structures, while the 2D cross-sectional views provide high-resolution images essential for precise boundary determination. This effective fusion addresses the challenges posed by complex, irregularly shaped gliomas, significantly reducing ambiguity and enhancing classification accuracy. A critical factor contributing to the method's superior performance is the utilization of ResNet50, a powerful pre-trained model known for its ability to capture both local and global spatial information. The architecture of ResNet50, with its deep layers and skip connections, allows it to learn complex features and patterns within the data, facilitating effective generalization across diverse glioma cases. This capability is particularly advantageous in medical imaging, where variability in tumor shapes and sizes can complicate analysis.

The dual-modality approach not only improves spatial analysis of glioma components, normal brain tissues, and critical anatomical structures, but it also enhances the overall classification performance. By combining 2D and 3D data, the proposed method significantly refines the representation of tumors, enabling a more detailed and accurate depiction essential for effective diagnosis and treatment planning.

In comparison to other techniques, the proposed fusion method excels in managing the irregular shapes and complex structures characteristic of gliomas. This comprehensive analysis not only bolsters classification accuracy but also yields invaluable insights into the spatial relationships between glioma components and surrounding tissues. These insights are crucial for optimizing treatment plans, ultimately leading to improved patient outcomes.

## Conclusion

Glioma segmentation and classification play a crucial role in ensuring timely diagnosis and effective treatment planning. The specialized architecture of UNET was particularly effective in delineating and segmenting regions of interest within images and made it suitable for glioma segmentation. UNET's architecture, with a contracting and expansive path, was well-suited for capturing spatial relationships within images. The proposed cascade deep learning model, incorporating UNET for Segmentation with a Dice score of 96.36% and an accuracy of 97.01%. ResNet50 for Classification achieves a remarkable accuracy of 99.25% for Subclass Classification (No Tumor, Necrotic or non-enhancing tumor, Peritumoral Edema, and Enhancing Tumor). During training, the model exhibits an average time of 176 s per epoch, and during validation, 13 s per epoch. The enhanced performance stems from the innovative fusion of segmented 2D and 3D data, coupled with the implementation of ResNet50 as a pretrained deep learning model for glioma classification, which is known for its ability to capture intricate features and patterns in data. Notably, the implemented model addresses challenges like the vanishing gradient issue by incorporating ResNet-50, leveraging skip connections to bypass layers. This 50-layered residual network effectively handles overfitting concerns during classification tasks subclass glioma images, utilizing both convolutional and identity blocks. The versatility of this cascade model extends its applicability to various other medical imaging tasks.

**Table 4** Performance evaluation with works from literature

| Research Work | Features extracted | Classification methods | Datasets | Accuracy | Precision | Sensitivity | Specificity |
|---|---|---|---|---|---|---|---|
| Prasetyo et al., 2022 [10] | High-level feature maps extracted from flattened layers/global average pooling layers | VGG16, VGG19, ResNet50, ResNet101, InceptionResNetV2 | BraTS 2019 | 99.00% | NA | NA | NA |
| Zahid et al., 2022 [5] | Differential equations and Swarm optimization algorithms, Single–fused feature vector, PCA | ResNet101 | BraTS 2018 | 94.40% | 96.75% | 96.75% | NA |
| Zheng et al. 2023 [3] | Unified Visualization and Classification Network: subregion-based attention mechanism, multiscale feature maps are fused to achieve higher resolution | UniVisNet | BraTS 2019 | 89.3% | NA | 90.4% | 85.3% |
| Sharif et al. [4] | Fusion of EKbHFV and MGA-based features | Cubic SVM classifier | BraTS 2018 and 2019 | 95% | 99.8% | 99.9% | NA |
| Bhatele et al. [6] | Combining Discrete Wavelet Decomposition, Central Pixel Neighbourhood Binary Pattern, and Gray Level Run Length Matrix | eXtreme Gradient Boosting with random forest classifier | BraTS 2013 and 2015 | 99.25% | NA | NA | NA |
| Ullah et al. [7] | T1C+T2+Flair MRI sequences ResNet50 and Stack Encoder-Decoder network parallel pooling layer fusion, Grey Wolf with updated criteria of the Jaya algorithm | Wide Neural Network classifier | BraTS 2020 and 2021 | 99.6% | 99.55% | 99.58% | NA |
| Ge et al. [8] | Multi-stream deep Convolutional Neural Network (CNN) architecture that extracts and fuses the features from multiple sensors | Softmax | BraTS 2017 | 90.87% | NA | NA | NA |
| Haq et al. [11] | Apply conditional random fields to refine outputs by incorporating spatial information for precise segmentation tasks | Deep CNNs | BraTS 2018 and Figshare | 96.5% | NA | NA | NA |
| Z. Huang et al. [12] | Modified activation function | Convolutional neural network based on complex networks (CNNBCN) | Figshare | 95.49% | NA | NA | NA |
| Jader et al. [13] | Texture based feature segmentation using GLCM and PCA | VGG-16, ResNet-50, and AlexNet | BraTS 2021 | 99.16% | NA | NA | NA |
| Aamir et al. [14] | Feature vectors—combined to form a hybrid feature vector using the partial least squares (PLS) | Agglomerative clustering with softmax | Figshare | 98.95% | NA | NA | NA |
| Proposed Work | Fusion of Segmented 2D and 3D MRI using weighted averaging | ResNet50 | BraTS 2018, 2019, 2020 | 99.25% | 99.32% | 99.10% | 99.77% |

Potential challenges in the proposed approach include the influence of dataset variety on model effectiveness, as variations in imaging protocols, equipment, and demographics can impact generalizability to new clinical environments. Incorporating data from multiple sources or developing a more diverse training set could enhance the model's adaptability, ensuring consistent performance across varied populations. Additionally, the use of fixed weighting in data fusion may limit the model's ability to adjust to unique dataset characteristics. Fixed weights assume equal or preset contributions from each input, which may be suboptimal in cases where different modalities offer complementary information. Dynamic or adaptive weighting could address this as future work, allowing the model to tailor its approach based on input specifics. Furthermore, as the model is scaled to larger, more diverse datasets, computational efficiency becomes a critical factor. High-dimensional, multi-modal data processing can require substantial resources, affecting real-time and large-scale application feasibility. Exploring computational optimizations, such as reducing model complexity without compromising accuracy, could help make this approach more practical for clinical use.

**Author contributions** Both authors contributed to the study conception and design. Material preparation, data collection was performed by Kiranmayee Janardhan and analysis was performed by Kiranmayee Janardhan and Christy Bobby Thomas.

**Funding** The authors did not receive support from any organization for the submitted work. No funding was received to assist with the preparation of this manuscript. No funding was received for conducting this study.

**Data availability** The datasets for the Multimodal Brain Tumor Segmentation Challenges are available on at https://www.med.upenn.edu/sbia/brats2018/data.html. https://doi.org/10.1109/TMI.2014.2377694.

## Declarations